\documentclass[showpacs,preprintnumbers,amsmath,amssymb]{revtex4}
\usepackage{graphicx}% Include figure files
\usepackage{dcolumn}% Align table columns on decimal point
\usepackage{bm}% bold math
\usepackage{graphicx}
\usepackage{amssymb}
\usepackage{amsmath}
\begin{document}

\title{On the smooth cross-over transition from the $\Delta$-string to
the Y-string three-quark potential}
\author{V. Dmitra\v sinovi\' c,\\
{\it Vin\v ca Institute of Nuclear Sciences, lab 010} \\
{\it P.O.Box 522, 11001 Beograd, Serbia} \\
Toru Sato,\\
{\it Dept. of Physics, Graduate School of Science, Osaka University}\\
{\it Toyonaka 560-0043, Japan} \\
Milovan {\v S}uvakov \\
{\it Institute of Physics,}\\
{\it Pregrevica 118, Zemun, P.O.Box 57, 11080 Beograd, Serbia,}}
\date{\today}
%
%======================
\begin{abstract}
We comment on the assertion made by Caselle {\it et al.}
\cite{Caselle:2005sf} that the confining (string) potential for
three quarks ``makes a smooth cross-over transition from the
$\Delta$-string to the Y-string configuration at interquark
distances of around 0.8 fm". We study the functional dependence of
the three-quark confining potentials due to a Y-string, and the
$\Delta$ string and show that they have different symmetries,
which lead to different constants of the motion (i.e. they belong
to different ``universality classes" in the parlance of the theory
of phase transitions). This means that there is no ``smooth
cross-over" between the two, when their string tensions are
identical, except at the vanishing hyper-radius. We also comment
on a certain two-body potential approximation to the Y-string
potential.
\end{abstract}

% 12.39.Pn Potential models
% 14.20.-c Baryons (including antiparticles)

\maketitle

%======================

%=====================
\section{Introduction}
%=====================

\begin{figure}[tbp]
\centerline{\includegraphics[width=2.5in,,keepaspectratio]{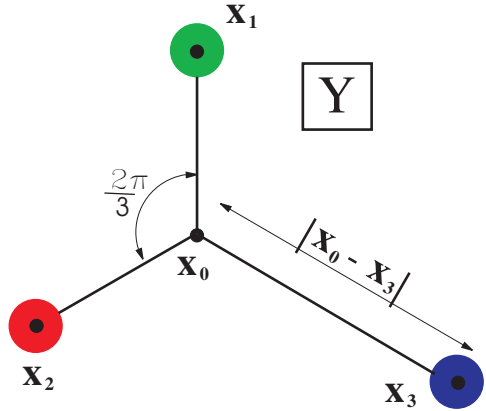}}
\caption{Three-quark Y-junction string potential.} \label{Y_3q}
\end{figure}

\begin{figure}[tbp]
\centerline{\includegraphics[width=2.5in,,keepaspectratio]{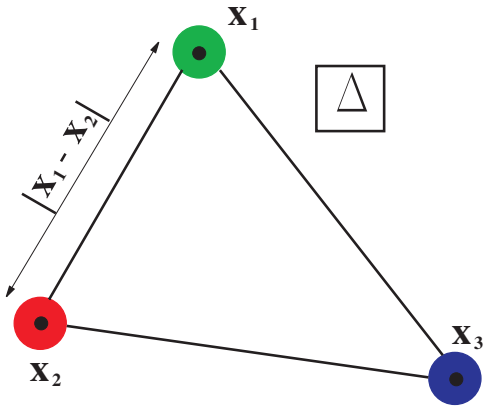}}
\caption{Three-quark $\Delta$-shape string potential.}
\label{D_3q}
\end{figure}

The so-called Y-junction string three-quark potential, defined by
\begin{equation}
\label{conf_Y} V_Y = \sigma \min_{\bf x_0}\; \sum_{i=1}^3 |{\bf
x_i} - {\bf x_0}|.
\end{equation}
has long been advertised \cite{artr75,dosc76} as the natural
approximation to the flux tube confinement mechanism, that is
allegedly active in QCD. Lattice investigations, Refs.
\cite{taka01,Alex01}, however, contradict each other in their
attempts to distinguish between the Y-string, Fig. \ref{Y_3q}, and
the $\Delta$-string potential, see Fig. \ref{D_3q},
\begin{equation}
\label{conf_D} V_{\Delta} = \sigma \sum_{i<j =1}^3 |{\bf x}_{i} -
{\bf x}_{j}|,
\end{equation}
which, in turn, is indistinguishable from the sum of three linear
two-body potentials. One may therefore view the present lattice
results as inconclusive and await the next generation of lattice
calculations\footnote{For more recent lattice results see Refs.
\cite{Bornyakov:2004uv}, \cite{Iida:2008cg}.}. Another point of
view held among some lattice QCD practitioners
\cite{deForcrand:2005vv} is that there should be a smooth
cross-over from the $\Delta$ to the Y-potential at interquark
distances of around 0.8 fm. This opinion is based on certain
similarities between the Potts model and lattice QCD which were
made more precise in Ref. \cite{Caselle:2005sf}. It was not stated
in Ref. \cite{Caselle:2005sf}, however, how exactly this
cross-over should be implemented, nor what they meant by
``interquark distances".

In the course of our studies of the (difference between the)
Y-string and the $\Delta$-string potentials \cite{dss09}, we have
taken this assertion at face value and tried to devise a ``smooth
cross-over", i.e. to make a smooth interpolation between
%function that smoothly connects
these two potentials, that is as simple as possible.
%and then solve the Schr\" odinger equation with the resulting potential.
What we found is simple enough to state: there can be no smooth
cross-over interpolation between
%function that connects
these two potentials; there must be always a discontinuity in some
variable(s). This fact may perhaps even be simply understood on
the basis of the different topologies of the two configurations,
see the discussion below, but the proof which we show (in some
detail) is complicated by the (technical) requirements of the
$S_3$ permutation symmetry.

We shall address the above questions one after another and for
this reason we divide the paper in four sections. In Sect.
\ref{sec1}, we define the potentials that we use, in the second
Sect. \ref{s:perm1}, we show an analytic proof of incompatibility
of $\Delta$ and Y-strings, and finally the third Sect.
\ref{s:approx}, addresses approximations to the string potentials
that can be used to extract results from the lattice or to be used
in the constituent quark model. The final Section \ref{s:summary}
contains a summary of our results and the discussion.

\section{Three-body potentials}
\label{sec1}

Any reasonable static, spin-independent three-body potential must
be: 1) translation-invariant, which means that it must depend only
on the two linearly independent relative coordinates, which we
call $({\bm \rho}, {\bm \lambda})$, but not on the center-of-mass
coordinate; 2) rotation-invariant, which means that it may depend
only on the three scalar products of the relative coordinates
${\bm \rho}^2, {\bm \lambda}^2, ({\bm \rho} \cdot {\bm \lambda})$;
and 3) permutation-invariant, which means that it may depend only
on certain combinations, yet to be determined, of the above three
scalar products of the relative coordinates.

We shall show that there are (precisely) three independent
permutation symmetric functions/variables of the relative
coordinates, that are related to simple geometrical/physical
properties (the moment of inertia, the area, and the perimeter of
the triangle) which clearly distinguishes them, and that the
potential's dependence on any one of them in particular carries
dynamical consequences.

Then we show that the (central, or three-body part of, for precise
definition see Sect. \ref{s:defY} below) Y-string potential
depends on only two (the moment of inertia and the area of the
triangle, but not on the perimeter) of these three variables in
most geometrical configurations; whereas the $\Delta$-string
potential depends only on the perimeter.

\subsection{Derivation of the $\Delta$- and Y-string potentials}
\subsubsection{Derivation of the $\Delta$-string potentials}
\label{s:defD}

The $\Delta$-string potential
\begin{equation}
\label{conf_D} V_{\Delta} = \sigma \sum_{i<j =1}^3 |{\bf x}_{i} -
{\bf x}_{j}| = 2 \sigma s,
\end{equation}
is proportional to the perimeter of the triangle $2 s = (a + b +
c)$, where $a = AB, b = BC, c = CA$ are the three sides of the
triangle and A,B,C are a positions of the quarks. When written in
this form, the potential is manifestly translation-, rotation- and
permutation invariant.

\subsubsection{Derivation of the Y-string potential} \label{s:defY}
\label{s:defY}

Three strings (``flux tubes") merge at the point $\bf x_{0}$,
which is chosen such that the sum of their lengths $l_{3q} =
l_{Y}$ is minimized
\begin{equation}
\label{conf_Y} l_{\rm Y} = \min_{\bf x_0}\; \sum_{i=1}^3 |{\bf
x_i} - {\bf x_{0}}|.
\end{equation}
If all the angles in the triangle are less than 120$^{\circ}$,
then the equilibrium Y-junction position is the so-called
Toricelli (or Fermat, or Steiner) point of classical geometry, $I
= {\bf x}_{0}$ in Fig.~\ref{Y_3q} that has the property that the
straight lines emanating from the junction point ${\bf x}_{0}$ and
leading to the quarks (``strings") all form an angle $2 \pi/3$ at
(see Fig. \ref{Y_3q}). The corresponding ``three-string length
$l_Y$ is
\begin{eqnarray}
l_Y &=& \sqrt{\frac{1}{2}(AB^2 + BC^2 + CA^2) + 2 \sqrt{3} {\rm
Area \triangle ABC}} \quad \text{if} \quad \measuredangle
\widehat{A}, \measuredangle \widehat{B}, \measuredangle
\widehat{C} \leq 120^{\circ} , \label{lYa1} \
\end{eqnarray}
where $a = AB, b = BC, c = CA$ are the three sides of the
triangle. Here one can see that the ``three-string" potential Eq.
(\ref{lYa1}) depends on the ``harmonic oscillator" variable $(a^2
+ b^2 + c^2)$, which is permutation symmetric and proportional to
the moment of inertia (divided by the quark mass) of this triangle
\footnote{The (trace of the tensor of) moment of inertia $I =
\sum_{i=1}^{3}\; m\;|{\bf x_i} - {\bf x_{\rm CM}}|^2 = m\;(a^2 +
b^2 + c^2) = 3 m\; R^2 = 3 m\; (\rho_{12}^{2} + \lambda_{12}^{2})
= 3 m (\rho _{23}^{2} + \lambda_{23}^{2}) = 3 m\; (\rho _{31}^{2}
+ \lambda _{31}^{2})$, where ${\bf x_{\rm CM}}$ is the radius
vector of the center-of-mass.}, and on the triangle area
\begin{eqnarray}
{\rm Area \triangle ABC} &=& \frac{1}{4} \sqrt{(AB + BC + CA)(-AB
+ BC + CA)(AB - BC + CA)(AB + BC - CA)} \nonumber \\
&=& \triangle (a, b, c) = \frac{1}{4} \sqrt{2 (a^2 b^2 + b^2 c^2 +
b^2 c^2) - a^4 - b^4 - c^4}, \label{Area1}
\end{eqnarray}
which is also permutation symmetric.

This form of the Y-string potential, when expressed in terms of
triangle sides $(a,b,c)$ only, exhibits its permutation symmetry
$S_3$, but hides the hidden/implicit angular dependence of the
potential, and potential interdependencies on other permutation
symmetric variables: The triangle area can be written using
Heron's formula
\begin{eqnarray}
{\rm Area \triangle ABC} &=& \sqrt{s(s - a)(s - b)(s - c)},
\label{Area2}
\end{eqnarray}
as a function of the triangle's semi-perimeter $s = \frac{1}{2}(a
+ b + c)$ and the three sides $a, b, c$.

It should be intuitively clear, however, that the perimeter and
the area of the triangle are two independent properties of the
triangle. Moreover, all three variables have non-zero
dimensionality, which seems to imply that there are three
different measures of ``intequark distances". The task now becomes
to find the most suitable independent coordinates/variables for an
interpolation.

If one of the angles within the triangle equals or exceeds the
value $2\pi/3=120^{\circ}$, however, the corresponding vertex of
the triangle is the junction point (although the Toricelli point
can be constructed in that case, as well, but generally lies
outside of the triangle and need not coincide with the vertex)
%When one of the angles equals $120^{\circ}$, the Fermat-Toricelli
%point coincides with the corresponding vertex of the triangle
and the minimal two-string length is
\begin{eqnarray}
l_V &=& (a + c) \quad \text{if} \quad \alpha = \measuredangle
\widehat{A} \geq 120^{\circ}, \quad \cos \alpha = \frac{1}{2 a c}
(a^2 + c^2 - b^2) \leq - \frac{1}{2} \\
&=& (a + b) \quad \text{if} \quad \beta = \measuredangle
\widehat{B} \geq 120^{\circ}, \quad \cos \beta = \frac{1}{2 a b}
(a^2 + b^2 - c^2) \leq - \frac{1}{2} \\
&=& (b + c) \quad \text{if} \quad \gamma = \measuredangle
\widehat{C} \geq 120^{\circ}, \quad \cos \gamma = \frac{1}{2 b c}
(b^2 + c^2 - a^2) \leq - \frac{1}{2} \\
&& \quad \text{where $a = AB, b = BC, c = CA$}. \label{lYa2}
\end{eqnarray}
Each one of these three expressions explicitly violates the
permutation symmetry (because of the ``missing piece of string"),
but taken in totality they maintain it, in the sense that no one
vertex is different from the others when its angle exceeds
$2\pi/3=120^{\circ}$.

\subsection{String potentials in terms of relative coordinates}
\label{s:jacobi}

The three sides of a triangle are not be the most useful
coordinates for practical calculations, however. Below, we shall
express these potentials in terms of conventional three-body
Jacobi relative coordinates ${\bm \rho},{\bm \lambda}$
\begin{figure}[tbp]
\centerline{\includegraphics[width=2.5in,,keepaspectratio]{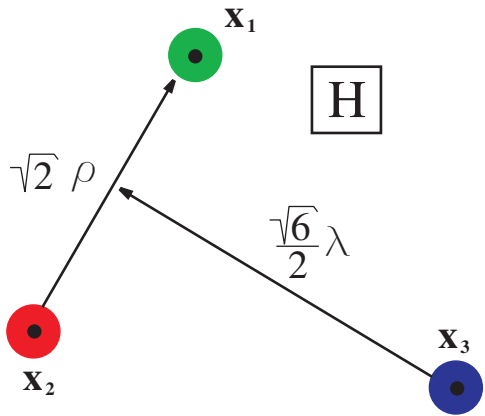}}
\caption{Two three-body Jacobi coordinates $\rho, \lambda$ that
define the ``hyper-radial" string length $\sqrt{|{\bm \rho}|^2 +
|{\bm \lambda}|^2}$.} \label{H_3q}
\end{figure}
\begin{eqnarray}
{\bm \rho}_{12} &=& \frac{1}{\sqrt{2}}({\bf x_1} - {\bf x_2}),
\label{e:rho} \\ %[1mm]
{\bm \lambda}_{12} &=& \frac{1}{\sqrt{6}}({\bf x_1} + {\bf
x_2}- 2 {\bf x_3}), \label{e:lambda} \\
{\bf x_{\rm CM}} &=& \frac{1}{3}({\bf x_1} + {\bf x_2} + {\bf
x_3}), \label{e:lambda}\
\end{eqnarray}
which obscures the permutation symmetry, however (see Sect.
\ref{s:perm}). From now on, we always specialize to the pair (12)
and drop the 12 index everywhere, so that the Jacobi coordinates
in our problem will be denoted by ${\bm \rho}$ (instead of ${\bm
\rho}_{12}$) and ${\bm \lambda}$ (instead of ${\bm
\lambda}_{12}$).

\subsubsection{String potentials in terms of hyper-spherical coordinates}

\begin{figure}[tbp]
\centerline{\includegraphics[width=2.5in,,keepaspectratio]{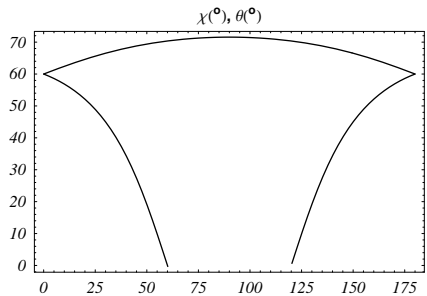}}
\caption{The boundary in the $\chi$ vs. $\theta$ plane, between
the regions in which the two- and the three-string potentials are
appropriate, see Eqs. (\ref{e:boundary}). The three-string
potential holds in the central region, whereas the two-string
potential holds in the ``corners" of this parallelogram.}
\label{f:bound0}
\end{figure}

We receive help here in the form of hyper-spherical coordinates:
instead of the moduli $\rho$ and $\lambda$ of the two Jacobi
vectors ${\bm \lambda}, {\bm \rho}$, shown in Fig. \ref{H_3q}, the
hyper-spherical coordinates introduce the hyper-radius,
%1. pre formule (15) dodati ... hyper-radius,
which is permutation symmetric:
%2. formule (17) nisam proverio
\begin{equation}
R = \sqrt{\rho^{2} + \lambda^{2}} ,
\end{equation}
as the only variable with dimension of length, the hyper-angle
$\chi$ through the polar transformation %(a stereographic
%projection of the sphere in six-dimensional space)
\begin{equation}
\rho = R \sin \chi,\quad \lambda = R \cos \chi \quad \text{with}
\quad 0 \le \chi \le \pi/2 .
\end{equation}
and the (physical) angle $\theta$ between ${\bm \lambda}$ and
${\bm \rho}$: $\cos\theta = {\bm \lambda}\cdot{\bm \rho}/(\rho
\lambda)$. The boundary in the $\chi$ vs. $\theta$ plane between
the regions in which the two- and the three-string potentials are
valid is determined by Eqs. (\ref{e:boundary}). There are three
such boundaries, determined by the three (in)equalities, that
merge continuously one into another at two ``contact points" and
one line, see Fig. \ref{f:bound0}.
\begin{eqnarray}
\cot\chi_1(\theta) &=& \frac{-1}{{\sqrt{3}}\,\cos\theta +
\sin\theta}, \nonumber \\
\cot\chi_2(\theta) &=& \frac{1}{{\sqrt{3}}\,\cos\theta -
\sin\theta},  \nonumber \\
\cot\chi_3(\theta) &=& \frac{1}{3}{\sqrt{5 - 2\,{\cos}^2\theta -
2\,|\sin\theta|\,{\sqrt{4 - \cos^2\theta}}}}. \label{e:boundary} \
\end{eqnarray}

The two hyper-angles ($\chi$, $\theta$) describe the shape of the
triangle, so the ($z=\cos2\chi$, $x=\cos\theta$) plane (square)
may be termed the ``shape-space". Manifestly, for each angle
$\theta$ there is another configuration with angle $\pi - \theta$
that describes a ``similar" triangle geometry that is a mirror
image of the other. For this reason any three-body potential
defined by geometric variables must be symmetric under reflections
across the $\theta = \pi/2$ axis. In Figs. \ref{f:delta contour},
\ref{f:Y contour} we show contour plots of the Y-string and the
$\Delta$-string potentials, respectively, in the ``shape space"
plane, i.e as functions of $z=\cos2\chi$ (vertical axis) and
$x=\cos\theta$ (horizontal axis), where this symmetry is plain to
behold.
\begin{figure}[tbp]
\centerline{\includegraphics[width=2.5in,,keepaspectratio]{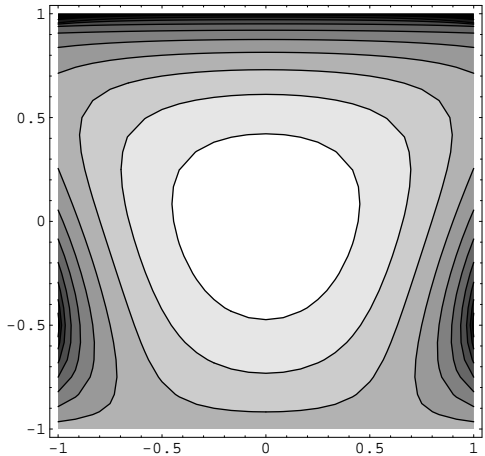}}
\caption{Contour plot of the $\Delta$-string potential as a
function of the cosines of the two hyper-angles $z=\cos2\chi$
(vertical axis) and $x=\cos\theta$ (horizontal axis) at any fixed
value of the hyper-radius $R$. The darker regions indicate a
smaller value of the potential. The reflection symmetry about the
x=0 axis should be visible to the naked eye.} \label{f:delta
contour}
\end{figure}
\begin{figure}[tbp]
\centerline{\includegraphics[width=2.5in,,keepaspectratio]{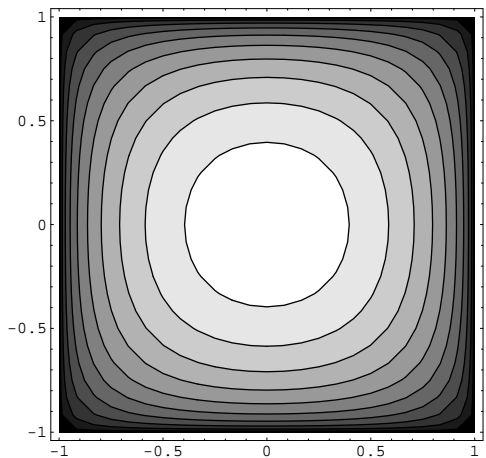}}
\caption{Contour plot of the Y-string potential as a function of
the cosines of the two hyper-angles $z=\cos2\chi$ (vertical axis)
and $x=\cos\theta$ (horizontal axis) at any fixed value of the
hyper-radius $R$. The darker regions indicate a smaller value of
the potential. Note, however, that this potential is not valid in
the whole region of ``shape space" depicted here: the ``two-body"
potential is valid beyond the boundary shown in Fig.
\ref{f:bound1}.} \label{f:Y contour}
\end{figure}
We see that the functional forms of these two potentials are
different: even their symmetries, obvious to the naked eye, are
different - one is symmetric only under reflections w.r.t vertical
axis, see Fig. \ref{f:delta contour}, whereas the other is also
symmetric under reflections w.r.t. the diagonals, as well as the
vertical and horizontal axes, see Fig. \ref{f:Y contour}. Below we
shall show that the Y-string potential has a continuous O(2)
dynamical symmetry, that is the source of the ``extra symmetry"
visible in Fig. \ref{f:Y contour}.

Due to their manifestly different symmetries, the two potentials
cannot coincide in the whole $(z,x)$ plane that describes the
shapes of the triangles. The intersection of the two potentials
may, at best, yield a curve in this plane of admissible triangle
configurations, which is but one real continuum $R$ out of a
double real continuum $R \times R$, which is measure-zero compared
with the disallowed configurations.

A simple numerical exercise, see Sect. \ref{s:diff1} below, shows
however that even that much does not happen, i.e. there is
absolutely no intersection of the two potentials when their string
tensions $\sigma_{\Delta} = \sigma_Y$ are equal, except at
vanishing hyper-radius $R=0$, i.e. when the triangle shrinks to a
point. That constitutes the proof of our contention that there is
no smooth transition from the $\Delta$ to the Y-string potential
at non-zero $R \neq 0$ and equal string tensions $\sigma_{\Delta}
= \sigma_{\rm Y}$.

\subsection{Difference of string potentials}
\label{s:diff}
\subsubsection{Difference of string potentials as a function of hyper-angles
at equal string tensions $\sigma_{\Delta} = \sigma_{\rm Y}$}
\label{s:diff1}

\begin{figure}[tbp]
\centerline{\includegraphics[width=2.5in,,keepaspectratio]{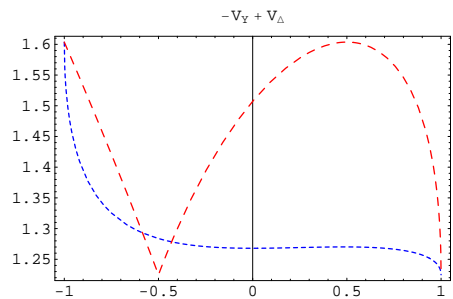}}
\caption{The difference of the $\Delta$ and the Y-string
potentials divided by the product of the string tension and the
hyper-radius $(V_{\Delta} - V_{\rm Y})/(\sigma R)$, as a function
of the cosine of the hyper-angle $z=\cos 2 \chi$, at fixed values
of $x=\cos \theta = \pm 1$ (red long dashes) and $x=\cos \theta =
0$ (blue short dashes). Note that the difference is always
(substantially) larger than unity, let alone zero.}
\label{f:diff1}
\end{figure}

The (normalized) difference of the $\Delta$ and the Y-string
potentials (with $\sigma R$ factored out) is shown in Fig.
\ref{f:diff1} as a function of the cosine of the hyper-angle $\cos
2 \chi$, at the fixed values of $\cos \theta = 0, \pm 1$. Note
that the difference is always (substantially) larger than zero,
i.e. that there are no zeros on the real line.
%4. u captionu slike 8. ...hyper-angle $z=cos 2 \chi$ (fali z=)
The same holds as a function of $\cos \theta$.
This shows that there is no continuous connection between the
$\Delta$ and the Y-string potentials at non-zero $R \neq 0$ and
equal string tensions $\sigma_{\Delta} = \sigma_{\rm Y}$.
%3. slika 7. i 8.: izraziti u jedinicama hyperradiusa ili jos bolje
%kao relativna razlika tj $(V_{\Delta}-V_Y)/V_{\Delta}$

One may try and change the string tensions $\sigma_{\Delta} \neq
\sigma_Y$, in which case there is an intersection of the two
surfaces, but that is not the problem that we started with. We
shall return to this new scenario below. Yet, even in that
scenario there can be a transition from one to another type of
string only in a limited sub-space of shapes.

\subsubsection{Difference of string potentials with unequal
string tensions $\sigma_{\Delta} \neq \sigma_{\rm Y}$}
\label{s:diff2}

\begin{figure}[tbp]
\centerline{\includegraphics[width=2.5in,,keepaspectratio]{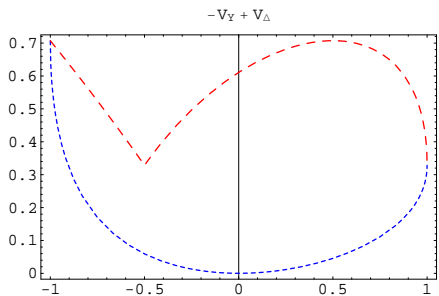}}
\caption{The difference of the $\Delta$ and the Y-string
potentials divided by the product of the string tension and the
hyper-radius $(V_{\Delta} - V_{\rm Y})/(\sigma R)$, as a function
of the cosine of the hyper-angle $z = \cos 2 \chi$, at two fixed
values of $x = \cos \theta = 0,1$ and $\sigma_{\rm
Y}/\sigma_{\Delta} = \sqrt{3} = 1.732$. Note that the difference
vanishes at only one point: $x=0=z$.} \label{f:diff2}
\end{figure}

\begin{figure}[tbp]
\centerline{\includegraphics[width=2.5in,,keepaspectratio]{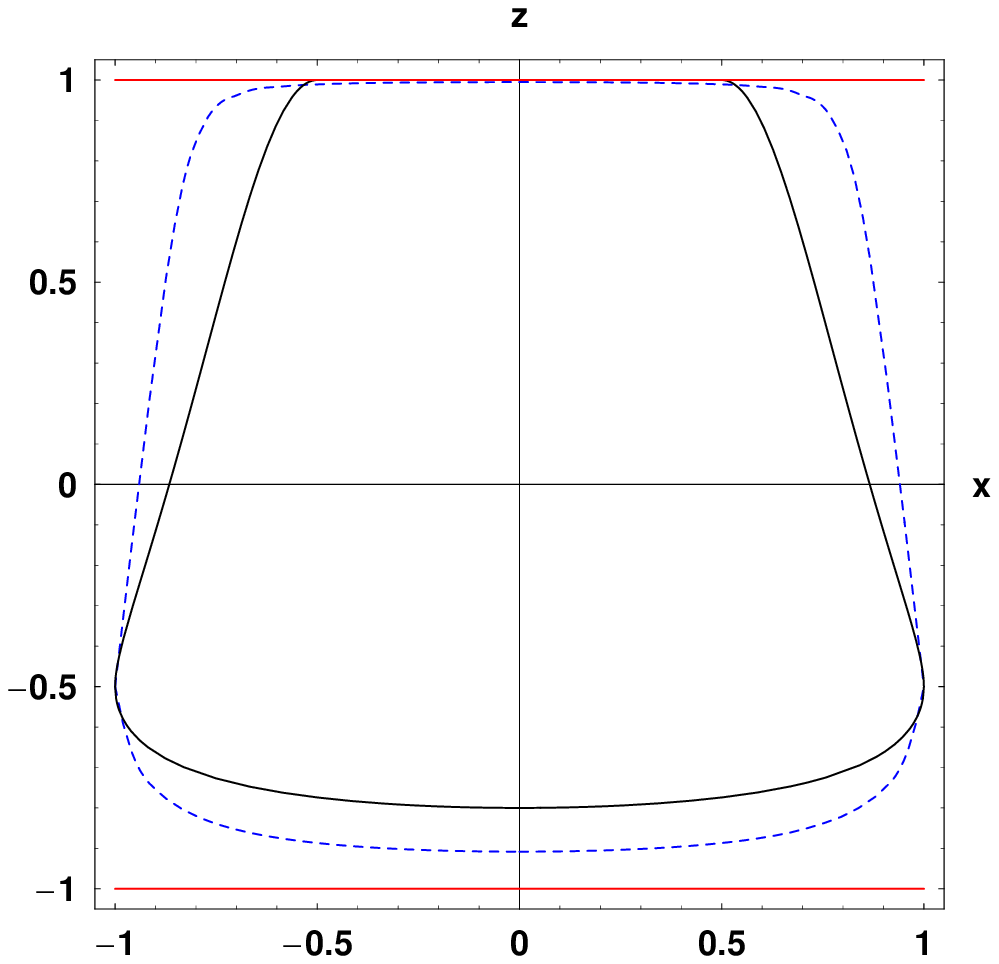}}
\caption{The locus of vanishing difference of the $\Delta$ and the
Y-string potentials (blue dashes) as a function of the cosine of
the hyper-angle $z=\cos 2 \chi$, and $x=\cos \theta$, at
$\sigma_{\rm Y}/\sigma_{\Delta} = 1.96$. Note that this whole
curve lies outside the range of validity (continuous black line)
of the three-body potential $V_{\rm Y}$, in which all of the
angles are smaller than $120^o$. } \label{f:diff3}
\end{figure}

One may readily circumvent the latter condition and then find some
small overlap of the two string potentials, see Figs.
\ref{f:diff2} and \ref{f:diff3}. Yet, even then the overlap region
is (only) a continuous curve in the $(x,z)$ plane, i.e. it covers
a negligibly small (``measure zero") ``number"/set of allowed
triangular configurations (``shapes") as compared with the %full
extent of all possible such sets, described by the complete
``shape space" $(x,z)$ plane. Note, moreover, that this overlap
lies within the ``allowed region" of shape space only in a limited
range of values of the ratio $\sigma_{\rm Y}/\sigma_{\Delta} \in
(\sqrt{3},1.86)=(1.732, 1.86)$. For higher values of this ratio
the overlap curve lies partially, or completely (for $\sigma_{\rm
Y}/\sigma_{\Delta} > 1.96$) outside the range of validity
(continuous black line in Fig. \ref{f:diff3}) of the three-body
potential $V_{\rm Y}$, so this solution is essentially irrelevant
\footnote{one should have rather compared with the two-body
potential Eqs. (\ref{lYa2}), which is not permutation symmetric,
however.}.
\begin{figure}[tbp]
\centerline{\includegraphics[width=2.5in,,keepaspectratio]{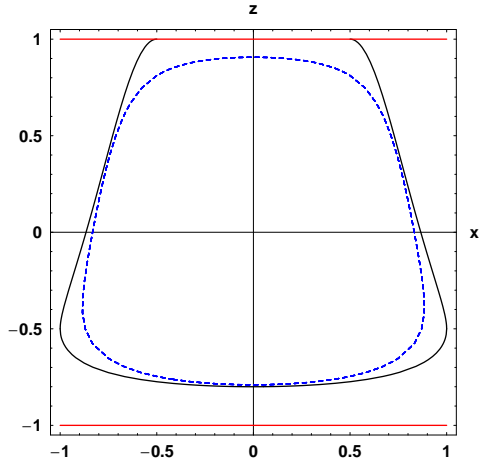}}
\caption{The locus of vanishing difference of the $\Delta$ and the
Y-string potentials (blue dashes) as a function of the cosine of
the hyper-angle $z=\cos 2 \chi$, and $x=\cos \theta$, at
$\sigma_{\rm Y}/\sigma_{\Delta} = 1.86$. Note that this whole
curve lies inside the range of validity (continuous black line) of
the three-body potential, in which all of the angles are smaller
than $120^o$. } \label{f:diff4}
\end{figure}

\section{Analytic proof of incompatibility of $\Delta$ and
Y-strings} \label{s:perm1}

This was just a numerical proof which did not expose the deeper
underlying reasons for this incompatibility: the (in)dependence of
these two strings on different permutation symmetric variables.
For an analytic proof of incompatibility we turn to the study of
the permutation symmetry in the three-body potential.

\subsubsection{Permutation symmetry properties of the Jacobi coordinates
and of the potential} \label{s:perm}

The Jacobi relative coordinate vectors $(\rho,\lambda)$ furnish a
two-dimensional irreducible representation of the $S_3$
permutation symmetry group. For example, let $P_{ij}$ be the
(two-body) $ij$-th particle permutation operator (transposition);
then
\begin{subequations}
\begin{eqnarray}
P_{12}\vec{\rho} & \rightarrow -\vec{\rho} \\
P_{12}\vec{\lambda} & \rightarrow \vec{\lambda} \\
P_{13}\vec{\rho} & \rightarrow \frac{1}{2}\vec{\rho} -
 \frac{\sqrt{3}}{2}\vec{\lambda} \\
P_{13}\vec{\lambda} & \rightarrow -\frac{\sqrt{3}}{2}\vec{\rho} -
 \frac{1}{2}\vec{\lambda}
\end{eqnarray}
\end{subequations}
Here we see that the $S_3$ permutation symmetry implies an
invariance (of the potential) under two rotations, through integer
multiples of $\frac{2 \pi}{3}=120^{\circ}$, and three reflections
in the $({\bm \rho}, {\bm \lambda})$ plane.\footnote{These
properties are perhaps most easily shown when the vectors are
expressed in terms of Simonov's \cite{Simonov:1965ei} complex
vector ${\bf z} = {\bm \lambda} + i {\bm \rho}$}. One special case
of such a permutation is the $P_{12}$ transposition, which is a
reflection that reverses the sign of $\cos \theta \to - \cos
\theta$, which is equivalent to the already mentioned
(geometrical) mirror symmetry.

Starting from $({\bm \rho},{\bm \lambda})$ one can construct one
symmetric ${\bf V}$, and two antisymmetric vectors: ${\bf A}, {\bf
{W}}$.
\begin{eqnarray}
{\bf V} &=& {\bm \lambda} \left({\bm \lambda}^2 - {\bm \rho}^2
\right) - 2 {\bm \rho} ({\bm \rho} \cdot {\bm \lambda})
\label{e:V}\\
{\bf A} &=& {\bm \rho} \left({\bm \lambda}^2 - {\bm \rho}^2
\right) + 2 {\bm \lambda} ({\bm \rho} \cdot {\bm \lambda})
\label{e:A}\\
{\bf {W}} &=& {\bm \rho} \times {\bm \lambda} \label{e:W}. \
\end{eqnarray}
The ``lengths" (norms) of vectors $V=|{\bf V}|$, $A=|{\bf A}|$,
$W=|{\bf W}|$ are invariant under the quark permutations. The
hyper-radius squared $R^2$ is the fourth permutation invariant
scalar. Of course, we expect only three out of four permutation
invariant scalars to be (non-linearly) independent. The non-linear
relationship reads
\begin{eqnarray}
V^2 + A^2 &=& R^2\left[R^4 - 4 W^2 \right] , \
\end{eqnarray}
so the third permutation symmetric variable may be taken as $(V^2
- A^2)$. Any (reasonable) confining three-body potential must be
permutation symmetric, so it must be a function of (only) $R, W$
and $(V^2 - A^2)$. As all three of these variables have non-zero
dimensions, one might think that each one represents a potentially
new definition of the ``interquark distance"; that is not the
case, as can be seen when one changes to hyper-spherical
coordinates.

\subsubsection{String potentials in terms of symmetrized
hyper-spherical coordinates}

The three permutation-symmetric hyper-spherical variables are $R$
and
\begin{eqnarray}
W &=& \frac{1}{2}\, R^2 \, \sin 2\chi \sin \theta \\
V^2 - A^2 &=& R^6\, \cos2\chi\, \left(\cos^2 2\chi - 3 \left(\sin
2\chi \,\cos \theta \right)^2 \right) \
\end{eqnarray}

When the interaction potential does not depend functionally on all
three permutation symmetric variables, then additional dynamical
symmetries appear: e.g. the ``central" part of the string
potential Eq. (\ref{lYa1}),
\begin{eqnarray}
V_{\rm Y} &=& \sigma \sqrt{\frac{3}{2} R^2 \left(1 + \sin 2 \chi
\sin \theta \right)} = \sigma \sqrt{\frac{3}{2} \left(R^2 + 2
W\right)}, \label{e:hypVY1} \
\end{eqnarray}
being a function of only $W$ and $R$, i.e. not a function of $(V^2
- A^2)$, is invariant under arbitrary rotations (not just through
integer multiples of $\frac{2 \pi}{3}=120^{\circ}$) in the $({\bm
\rho}, {\bm \lambda})$ plane. That leads to a new integral of
motion $G = {\bm \lambda} \cdot {\bf p_{\rho}} - {\bm \rho} \cdot
{\bf p_{\lambda}}$, associated with the dynamical symmetry (Lie)
group $O(2)$, see Sect. below, which is not conserved in the case
of the $\Delta$ string potential. Further, when the potential
depends only on $R$, the dynamical symmetry is extended to the
$O(6)$ Lie group.

The $\Delta$-string potential
\begin{eqnarray}
V_{\Delta} &=& \sigma \sum_{i<j =1}^3 |{\bf x}_{i} - {\bf
x}_{j}| \nonumber \\
&=& \sigma \left(\sqrt{2} \rho + \sqrt{\frac32}\left|{\bm \lambda}
+ \frac{{\bm \rho}}{\sqrt{3}}\right| + \sqrt{\frac32}\left|{\bm
\lambda} - \frac{{\bm \rho}}{\sqrt{3}}\right|\right)
\label{e:hypVY2} \
\end{eqnarray}
can be expressed in terms of $({\bf V}, {\bf A})$: just solve Eqs.
(\ref{e:V}),(\ref{e:A}) for $({\bm \rho}, {\bm \lambda})$ as
functions of $({\bf V}, {\bf A})$ and insert the results
\begin{eqnarray}
{\bm \rho} &=& \left({\bf V} \left({\bm \lambda}^2 - {\bm \rho}^2
\right) + 2 {\bf A} ({\bm \rho} \cdot {\bm \lambda})\right)
\left(R^2 - \left(2 W \right)^2 \right)^{-2}
\label{e:rho}\\
{\bm \lambda} &=& \left({\bf A} \left({\bm \lambda}^2 - {\bm
\rho}^2 \right) - 2 {\bf V} ({\bm \rho} \cdot {\bm
\lambda})\right) \left(R^2 - \left(2 W \right)^2 \right)^{-2}
\label{e:lambda}\
\end{eqnarray}
i.e.
\begin{eqnarray}
{\bm \rho} &=& \left({\bf V} \cos2\chi  + {\bf A} \left(\sin 2\chi
\,\cos \theta \right)\right) \left(1 - \left(\frac{2 W}{R^2}
\right)^2 \right)^{-2}
\label{e:rho1}\\
{\bm \lambda} &=& \left({\bf A} \cos2\chi - {\bf V} \left(\sin
2\chi \,\cos \theta \right)\right) \left(1 - \left(\frac{2 W}{R^2}
\right)^2 \right)^{-2} \label{e:lambda1}\
\end{eqnarray}
into Eq. (\ref{e:hypVY2}) which shows that
\begin{eqnarray}
V_{\Delta} &=& V_{\Delta}(R,W,(V^2 - A^2)) \label{e:hypVY3} \
\end{eqnarray}
is a function of all three permutation symmetric three-body
variables, however. So, we see that in general the two string
potentials depend on different symmetric three-body variables, and
thus cannot be smoothly connected, except perhaps in special
cases/geometries, where the variable $(V^2 - A^2)$, and/or other
variable(s) vanish.

\subsubsection{Admissible region(s) for a ``smooth crossover" of two strings}
\label{s:admiss}

\begin{figure}[tbp]
\centerline{\includegraphics[width=2.5in,,keepaspectratio]{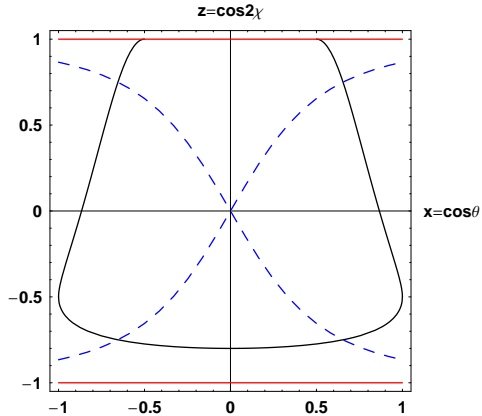}}
\caption{The boundary in the $\chi$ vs. $\theta$ plane (solid
line), between the regions in which one of the angles is larger
than $120^o$, and the curves along which the functional identity
of the $\Delta$ and the Y-string potentials is admissible (the
blue long-dashed curve).} \label{f:bound1}
\end{figure}

Thus, we need to solve
\begin{eqnarray}
V^2 - A^2 &=& R^6\, \cos2\chi\, \left(\cos^2 2\chi - 3 \left(\sin
2\chi \,\cos \theta \right)^2 \right) = 0. \label{e:va} \
\end{eqnarray}
There are two ``trivial" solutions: 1) $R=0$; 2) $\cos2\chi = 0$;
and a family of non-trivial solutions to 3) $\cos^2 2\chi - 3
\left(\sin 2\chi \,\cos \theta \right)^2 =0$, i.e. $\cot 2\chi =
\pm \sqrt{3} \,|\cos \theta|$, which determines the admissible
``smooth crossover" region (the blue long-dashed curve) in Fig.
\ref{f:bound1}.
Thus, we can see that out of the double-continuum (``plane") of
possible geometric configurations at a given ``interquark
distance" $R$, a ``smooth crossover" is potentially admissible,
though not guaranteed, only on a single continuum (the blue dashed
curves and the vertical axis at $x=0$ in Fig. \ref{f:bound1}).
Now, to make this an actual ``smooth crossover" region, rather
than merely an admissible one, one must find the intersection of
this curve and the locus of zeros of $V_{\Delta}-V_{\rm Y}$, that
we searched for in Sect. \ref{s:diff1}, to no avail.

This fact, together with our numerical results from Sect.
\ref{s:diff1}, are sufficient proof of our claim that even along
this curve the crossover is {\it never} smooth. This means that,
in order to provide a continuous transition from the $\Delta$ to
the Y-string configuration at any finite-$R$ crossover point,
including the $R = 0.8 fm$ one, the string tension $\sigma$ must
be variable and have a discontinuity in at least one (permutation
symmetric) variable. All-in-all, this shows that with constant
string tension $\sigma_{\rm Y}=\sigma_{\Delta}$ there can be no
smooth transition from the $\Delta$ to the Y-string configuration
in any geometry. Even when $\sigma_{\rm Y}/\sigma_{\Delta} \in
(1.732, 1.86)$ there are only three geometric configurations where
a smooth transition can be accomplished, see the discussion below
and Fig. \ref{f:bound2}.

\subsubsection{Dynamical symmetry of the Y-string potential}
\label{s:dyn symm}

As stated above, independence of the variable $(V^2 - A^2)$ leads
to the invariance under ``generalized rotation"
\begin{eqnarray}
\delta {\bm \rho} &=& \varepsilon {\bm \lambda} \nonumber \\
\delta {\bm \lambda} &=& - \varepsilon {\bm \rho} . \label{e:O2
trf} \
\end{eqnarray}
in the six-dimensional hyper-space and thus leads to the new
integral-of-motion $G = {\bm \lambda} \cdot {\bf p_{\rho}} - {\bm
\rho} \cdot {\bf p_{\lambda}}$, associated with the dynamical
symmetry (Lie) group $O(2)$ that is a subgroup of the full $O(6)$
Lie group. In certain cases the new integral of motion $G$ can be
integrated and the resulting holonomic constraint can be used to
eliminate one degree of freedom \cite{SD07}. In the case of the
$\Delta$ string potential this $G$ is not an integral-of-motion,
i.e. it is not constant in time, however, due to the absence of
the O(2) symmetry of this potential \cite{SD07}. The dynamical
O(2) transformation rules Eqs. (\ref{e:O2 trf}) can be applied to
the coordinates $(z,x)$ as follows:
\begin{eqnarray}
\delta x &=& 2 \varepsilon (1 - x^2)\frac{z}{\sqrt{1 - z^2}} \nonumber \\
\delta z &=& - 2 \varepsilon x \sqrt{1 - z^2} . \label{e:xz O2
trf} \
\end{eqnarray}
Note that these equations are non-linear in $(x,z)$ implying a
non-trivial transformation of the ``shape space" under this new
dynamical O(2) symmetry. Only near the origin $(x,z)=(0,0)$ does
this transformation look like an infinitesimal rotation; at the
edges $(x,z)=(\pm 1,\pm 1)$ it either vanishes, or diverges. One
can find another set of scalar variables $(x\sqrt{1-z^2},z)$ in
which the dynamical O(2) symmetry transformation rules Eqs.
(\ref{e:O2 trf}) are linear, but at the price of introducing
imaginary parts of the potential, at least in some regions of the
``shape space". A detailed study of this symmetry is a task for
the future \cite{SD07}.

%================
\section{Approximations to the string potentials}
\label{s:approx}
%================

\subsection{String approximations to the lattice:
the composite string}
\label{s:composite}

\begin{figure}[tbp]
\centerline{\includegraphics[width=1.5in,]{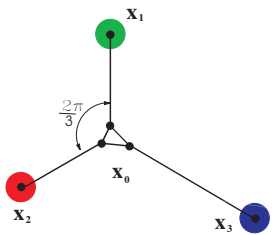}}
\caption{The new three-quark Y-junction string potential.}
\label{f:New Y_3q}
\end{figure}

One possible way to fit the lattice results is, perhaps, to have a
``composite Y-string" that contains a ``core triangle" of variable
size (proportional to, yet smaller than the quark triangle)
instead of the Y-junction point, see Fig. \ref{f:New Y_3q}. In
that case, however, one may not talk about a definite cross-over
inter-quark distance/hyper-radius, and one would still have
contributions from both kinds of string at all distances.
Note, however, that this prescription is precisely equivalent to
the sum of two strings with unequal tensions: for a given (fixed)
value of parameter $\alpha \in (0,1)$ (coefficient of
proportionality) one has
\begin{equation}
\label{composite} V_{\rm composite} = \alpha V_{\rm Y} + (1 -
\alpha) V_{\Delta} = \sigma_{\rm Y} \min_{\bf x_0} \sum_{i=1}^3
|{\bf x_i} - {\bf x_0}| +  \sigma_{\Delta} \sum_{i<j}^3 |{\bf
x}_{i} - {\bf x}_{j}|.
\end{equation}
There is at least one good reason for the unequal string tensions,
{\it viz.} the different color factors associated with the
quadratic and the cubic Casimir operators of SU(3), c.f. Ref.
\cite{vd01}. It appears ``natural" to associate the cubic Casimir
color factor with the Y-string and the quadratic Casimir color
factor with the $\Delta$ string. As there is no (reliable) way of
dealing with color non-singlets on the lattice (see, however the
work of Saito et al. \cite{saito05}) heretofore there was little
hope of extracting these factors (or at least their ratio) from
the lattice. This Ansatz allows, at least in principle, the
extraction of the ratio of the Y-string and the $\Delta$-string
tensions $\sigma_{\rm Y}/\sigma_{\Delta}$.

\begin{figure}[tbp]
\centerline{\includegraphics[width=2.5in,,keepaspectratio]{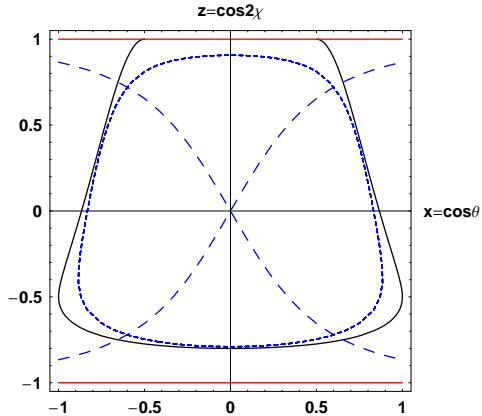}}
\caption{The boundary in the $\chi$ vs. $\theta$ plane (solid
line), between the regions in which all of the angles are smaller
than $120^o$ and the ones where they are not, the locus of zeros
of $V_{\Delta}-V_{\rm Y}$ at $\sigma_{\rm Y}/\sigma_{\Delta} =
1.86$ (the blue short-dashed curve), and the lines along which
identity of the $\Delta$ and the Y-string potentials is admissible
(the blue long-dashed curve).} \label{f:bound2}
\end{figure}

Note that in this case the smooth crossover from the $\Delta$ to
the Y-string is not entirely out of the question, albeit it is
still severely restricted in the shape space. The intersection of
the curve defined by Eq. (\ref{e:va}) and the locus of zeros of
$V_{\Delta}-V_{\rm Y}$, Figs. \ref{f:diff3}, \ref{f:diff4}, yields
the allowed crossover points/configurations, of which there are at
most six %of which only three are admissible
(out of a double-continuum, see Fig. \ref{f:bound2}), and that
only when $\sigma_{\rm Y}/\sigma_{\Delta} \in (1.732, 1.86)$. One
obtains a region of admissible values of $\alpha$ from the
admissible values of $\sigma_{\rm Y} = \alpha \; \sigma$ and
$\sigma_{\Delta} = (1 - \alpha) \; \sigma$. Then $\sigma_{\rm
Y}/\sigma_{\Delta} = \alpha/(1 - \alpha) \in (1.732, 1.86)$,
implies $\alpha \in (0.634, 0.650)$. If this turns out too
restrictive for the actual lattice results, then one may even
introduce a hyper-radially dependent coefficient $\alpha(R)$ that
peaks at some non-zero $R$, i.e. a not-quite-linearly rising
two-body confining potential.

In order to facilitate this separation of the Y-string from the
$\Delta$-string, the hyper-angular dependence of the three-body
potential ought to be expressed in terms of the new variables
$x^{'} = x\, \sqrt{1 - z^2}$ and $z^{'} = z$. Then the Y-string
component is manifested through the sole dependence on $\rho^2 =
\, \left(x^{'2} + \, z^{'2} \right) = \,(V^2 + A^2)R^{-6} = \,
\left(1 - \,\left(\frac{2W}{R^{2}}\right)^{2} \right)$, whereas
the $\Delta$-string is manifested through the dependence of the
potential on the new hyperangle $\phi = \tan^{-1}
\left(\frac{x^{'}}{z^{'}}\right)$, within the confines of the
``central potential" boundary (defined by Eqs. (\ref{e:boundary})
in terms of ``old" variables $(\chi,\theta)$).
%$(V^2 -A^2)R^{-6} = \, z^{'}\, \left(z^{'2} - 3\, x^{'2} \right)$.
If both the Y-string and the $\Delta$-string components are
present in the lattice three-body potential, their ratio can be
disentangled by measuring the ratio of the $\rho = \, \sqrt{x^{'2}
+ \, z^{'2}}$ to the $\phi = \tan^{-1}
\left(\frac{x^{'}}{z^{'}}\right)$ dependencies, again within the
confines of the ``central potential" boundary, because the
$\Delta$-string also depends on $\rho$.

Before closing this subsection, we ought to point out that
another, perhaps similar in spirit, way to fit the lattice results
has been devised and applied in Ref. \cite{taka01}. In that
``generalized Y-string Ansatz", a ``core circle" of a definite
radius has been assumed around the Y-junction point, see Fig. 14
in Ref. \cite{taka01}. This Ansatz is rather difficult to describe
analytically in terms of hyper-spherical coordinates, so its
validity would be difficult to ascertain on the lattice. The
``composite string" Ansatz, on the other hand, can be readily
recognized by its dependence on the ``third variable" which may be
chosen either as $(V^2 - A^2)\, R^{-6} = \, z\, \left(z^2 - 3\,
x^2\, \left(1 - z^2\right) \right) = R^6 \, z^{'} \, \left(z^{'2}
- 3\, x^{'2}\right)$, or as $\phi$.

\subsection{Optimal two-body approximation to the ``central" Y-string potential}
\label{s:approx Y}

In this light one may try using a linear combination of the
Y-string potential and a (linearly) rising two-body potential,
perhaps with a variable string tension in the constituent quark
model calculations. The  Y-string potential has been known for the
difficulty of implementation in the Schr\" odinger equation, which
has only recently been solved systematically, see
\cite{dss09},\cite{Narodetskii:2008uc}, so the authors of Refs.
\cite{fabr97},\cite{Grenoble03} tried to approximate the
``central" Y-string potential $V_{\rm Y}$ with a two-body
$V_{\Delta}$ plus possibly a one-body potential $V_{\rm CM}$, that
are easier to deal with numerically. Such approximations may still
be valuable in calculations of multi-quark states. Here we show
that one particular linear combination of $V_{\Delta}$ and $V_{\rm
CM}$
\begin{eqnarray}
V_{\rm comb.} &=& \frac12 \left(V_{\rm CM} + \frac{1}{\sqrt 3}
V_{\Delta} \right) = \frac{\sigma}{2}  \left(\sum_{i=1}^3 |{\bf
x_i} - {\bf x_{\rm CM}}| + \frac{1}{\sqrt 3}
\sum_{i<j}^3 |{\bf x}_{i} - {\bf x}_{j}| \right),
\nonumber \\
&=& \sigma \left(\sqrt{\frac23} (\rho + \lambda ) + \sqrt{\frac12}
\left(\left|{\bm \lambda} + \frac{{\bm \rho}}{\sqrt{3}}\right| +
\left|{\bm \lambda} - \frac{{\bm \rho}}{\sqrt{3}}\right| +
\left|{\bm \rho} + \frac{{\bm \lambda}}{\sqrt{3}}\right| +
\left|{\bm \rho} - \frac{{\bm
\lambda}}{\sqrt{3}}\right|\right)\right) \label{e:approx}  \
\end{eqnarray}
assumes the highest degree of dynamical symmetry possible with
these variables and a linear hyper-radial dependence, viz. the
symmetry under the exchange ${\bm \rho} \leftrightarrow {\bm
\lambda}$, that exceeds the usual $S_3$ permutation symmetry.

\begin{figure}[tbp]
\centerline{\includegraphics[width=2.5in,,keepaspectratio]{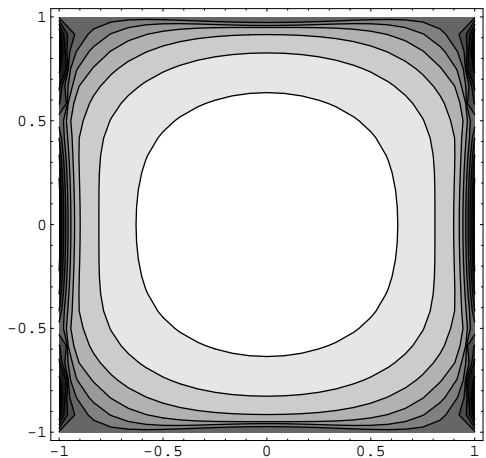}}
%Combined_contour.eps}}
\caption{Contour plot of the difference between the genuine
Y-string potential $V_{\rm Y}$, shown in Fig. \ref{f:Y contour},
and the combined CM- and $\Delta$-string potentials $\frac12
\left[V_{\rm CM} + 1/{\sqrt 3} V_{\Delta}\right]$ as a function of
the cosine of the hyper-angle $z=\cos 2 \chi$, and $x=\cos
\theta$. Note that the overall (``quasi-rotational") symmetry of
this figure is the same as that of Fig. \ref{f:Y contour}, except
in the corners and near the edges of the square, where the
difference is the largest. The darker regions indicate a larger
value of the difference of the potentials.}
\label{f:Combined_contour}
\end{figure}

That new symmetry does not amount to the exact O(2) symmetry of
the ``central" Y-string potential $V_{\rm Y}$, as yet, but is
numerically sufficiently close to it in the region of
applicability, see Fig. \ref{f:Combined_contour}, for most
practical purposes, like that of the quark model calculations.
In Fig. \ref{f:Russ Combined_contour}, we show the difference
between the combined CM- and $\Delta$-string potential $\frac12
\left[V_{\rm CM} + V_{\Delta}\right]$ used in Ref.
\cite{Grenoble03}. Note the conspicuous broadening of the contour
in the lower half-plane (the pear-shape of the contours) and
consequently the absence of the continuous ``quasi-rotational"
symmetry in this figure, which goes to show that this
approximation to the genuine Y-string potential does not have the
correct symmetry, which in turn is a consequence of the missing
factor $1/{\sqrt 3}$.
\begin{figure}[tbp]
\centerline{\includegraphics[width=2.5in,,keepaspectratio]{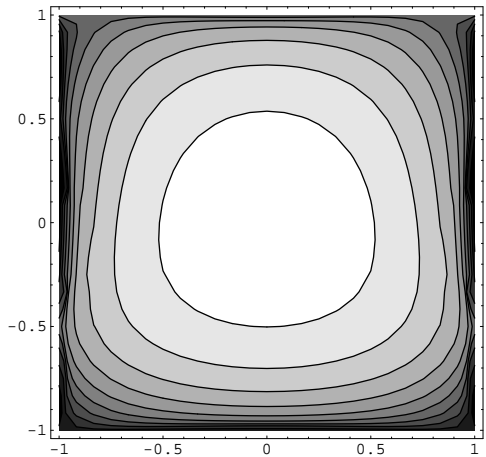}}
%Combined_contour.eps}}
\caption{Contour plot of the difference between the genuine
Y-string potential $V_{\rm Y}$ and the combined CM- and
$\Delta$-string potential $\frac12 \left[V_{\rm CM} +
V_{\Delta}\right]$ used in Ref. \cite{Grenoble03} as a function of
the cosine of the hyper-angle $z=\cos 2 \chi$, and $x=\cos
\theta$.  Note that the overall (``quasi-rotational") symmetry of
this figure is broken as compared with that of Fig. \ref{f:Y
contour}, due to the missing factor $1/{\sqrt 3}$. The difference
is the largest in the darkest regions.} \label{f:Russ
Combined_contour}
\end{figure}
%===============================================================

%One such combination was used in Ref. \cite{carl83}, albeit only
%in first order perturbation theory: The good agreement with the
%experimental value of the Roper mass was presumably achieved only
%by the predominance of the Coulomb interaction over the Y-string
%potential. Precise examination of such scenarios remains a task
%for the future.

%===============================================================
%================
\section{Summary and Discussion}
% and Outlook}
\label{s:summary}
%================

In summary, we have studied the functional dependence of the
three-quark confining potential due to a Y-string, and the
$\Delta$-string. We have found fundamentally different results for
these two kinds of strings, which lead to different constants of
the motion (``universality classes"). This means that there can be
no smooth crossover between the two, except at the vanishing
hyper-radius $R$, which is physically singular and mathematically
not very meaningful. Perhaps, this should be no surprise, as the
two strings have different topologies: one separates the plane
into two disjoint parts, whereas the other one does not.

\section*{Acknowledgments}
\label{ack}

We wish to thank the referee of our previous paper, Ref.
\cite{dss09} for drawing our attention to Refs.
\cite{Caselle:2005sf} and \cite{deForcrand:2005vv}. One of us
(V.D.) wishes to thank Prof. S. Fajfer for her hospitality at the
Institute Jo\v zef Stefan, Ljubljana, where this work was started
and to Profs. H. Toki and A. Hosaka for their hospitality at RCNP,
Osaka University, where it was continued.


\begin{thebibliography}{10}

\bibitem{artr75} X. Artru, Nucl. Phys. {\bf B 85}, 442 (1975).

\bibitem{dosc76} H. G. Dosch and V. Mueller, Nucl. Phys. {\bf
B 116}, 470 (1976).

\bibitem{taka01} T. T. Takahashi, H. Matsufuru, Y. Nemoto, and H.
Suganuma, Phys. Rev. Lett. {\bf 86}, 18 (2001); Phys. Rev D {\bf
65}, 114509 (2002).

\bibitem{Alex01}
C. Alexandrou, P. De Forcrand, A. Tsapalis, Phys. Rev. {\bf D 65},
054503,(2002).

\bibitem{Iida:2008cg}
  H.~Iida, N.~Sakumichi and H.~Suganuma,
  arXiv:0810.1115 [hep-lat].

%\cite{Bornyakov:2004uv},\cite{Iida:2008cg}
\bibitem{Bornyakov:2004uv}
  V.~G.~Bornyakov {\it et al.}  [DIK Collaboration],
  %``Baryonic flux in quenched and two-flavor dynamical QCD,''
  Phys.\ Rev.\  D {\bf 70}, 054506 (2004)
  [arXiv:hep-lat/0401026].
  %%CITATION = PHRVA,D70,054506;%%

\bibitem{Caselle:2005sf}
  M.~Caselle, G.~Delfino, P.~Grinza, O.~Jahn and N.~Magnoli,
  %``Potts correlators and the static three-quark potential,''
  J.\ Stat.\ Mech.\  {\bf 0603}, P008 (2006)
  [arXiv:hep-th/0511168].
  %%CITATION = JSTAT,0603,P008;%%

%\cite{deForcrand:2005vv}
\bibitem{deForcrand:2005vv}
  Ph.~de Forcrand and O.~Jahn,
  %``The baryon static potential from lattice QCD,''
  Nucl.\ Phys.\  A {\bf 755}, 475 (2005)
  [arXiv:hep-ph/0502039].
  %%CITATION = NUPHA,A755,475;%%

\bibitem{dss09}
  V.~Dmitra\v sinovi\a' c, T.~Sato and M.~Suvakov,
  Eur. Phys. J. {\bf C 62}, 383-398 (2009);
see also V. Dmitra\v sinovi\a' c, T. Sato and M. {\v S}uvakov,
``Low-lying states in the Y-string three-quark potential", p. 30 -
35, Proceedings of the Mini-Workshop ``Few-Quark states and the
Continuum'', ``Bled Workshops in Physics'', Vol. 9, No. 9. ed. B.
Golli, M. Rosina and S. \v Sirca, DMFA - Zalo\v zni\v stvo,
Ljubljana, Slovenia, (2008).

\bibitem{Narodetskii:2008uc}
  I.~M.~Narodetskii, C.~Semay and A.~I.~Veselov, Eur.\ Phys.\ J.\
  C {\bf 55}, 403 (2008)
  % [arXiv:0801.4270 [hep-ph]].

\bibitem{Simonov:1965ei}
Yu.~A.~Simonov, Sov.\ J.\ Nucl.\ Phys.\  {\bf 3}, 461 (1966)
[Yad.\ Fiz.\ {\bf 3}, 630 (1966)].

\bibitem{vd01}
V. Dmitra\v sinovi\' c, Phys. Lett. {\bf B 499}, 135 (2001); Phys.
Rev. {\bf D 67}, 114007 (2003).

\bibitem{saito05} %{\it QCD color interactions between two quarks}
A. Nakamura, T. Saito, Phys. Lett. {\bf B 621}, 171 (2005).

\bibitem{fabr97} M. Fabre de la Ripelle and M. Lassaut, Few-Body Systems
{\bf 23}, 75 (1997).

\bibitem{Grenoble03}
B. Silvestre-Brac, C. Semay, I. M. Narodetskii, and A.I. Veselov,
Eur.Phys.J. {\bf C32}, 385 (2003). I.M. Narodetskii and M.A.
Trusov, hep-ph/0307131v1.

\bibitem{SD07}
V. Dmitra\v sinovi\' c and M. \v Suvakov,
%{\it On the Y-string potential and its nonrelativistic classical motions},
in preparation (2009).

\end{thebibliography}
\end{document}